\documentclass[3p,final,11pt,authoryear]{elsarticle} 
\usepackage[english]{babel}
\usepackage[utf8]{inputenc}
\usepackage[T1]{fontenc}
\usepackage{lmodern}

\usepackage{graphicx,xcolor}
\usepackage{setspace}
\usepackage{subcaption,caption}
\usepackage{booktabs,tabu}
\usepackage{tabularx}
\usepackage{lineno}
\usepackage{times}

\usepackage{amsmath}
\usepackage{siunitx}

\usepackage{hyperref}
\hypersetup{colorlinks=true,linkcolor=blue}

\usepackage[nameinlink,noabbrev]{cleveref} 
\creflabelformat{equation}{#2#1#3}

\usepackage[acronym]{glossaries} 
\glsdisablehyper

\makeatletter
\def\ps@pprintTitle{%
  \let\@oddhead\@empty
  \let\@evenhead\@empty
  \let\@oddfoot\@empty
  \let\@evenfoot\@oddfoot
}
\makeatother

\input{macros.tex}

\begin{document}
	
	\begin{frontmatter}
		\title{ResLogit: A residual neural network logit model for data-driven choice modelling}
		\author[1]{Melvin Wong}\corref{cor1}
		\ead{melvin.wong@epfl.ch}
		
		\author[2]{Bilal Farooq}
		\ead{bilal.farooq@ryerson.ca}
		
		\address[1]{École Polytechnique Fédérale de Lausanne (EPFL), School of Architecture, Civil and Environmental Engineering (ENAC), Transport and Mobility Laboratory, Switzerland}
		\address[2]{Ryerson University, Laboratory of Innovations in Transportation (LiTrans), Canada}
		\cortext[cor1]{Corresponding Author.}
		
		\begin{abstract}
			This paper presents a novel deep learning-based travel behaviour choice model.
			Our proposed Residual Logit (ResLogit) model formulation seamlessly integrates a Deep Neural Network (DNN) architecture into a multinomial logit model. 
			Recently, DNN models such as the Multi-layer Perceptron (MLP) and the Recurrent Neural Network (RNN) have shown remarkable success in modelling complex and noisy behavioural data. 
			However, econometric studies have argued that machine learning techniques are a `black-box' and difficult to interpret for use in the choice analysis.
			We develop a data-driven choice model that extends the systematic utility function to incorporate non-linear cross-effects using a series of residual layers and using skipped connections to handle model identifiability in estimating a large number of parameters.
			The model structure accounts for cross-effects and choice heterogeneity arising from substitution, interactions with non-chosen alternatives and other effects in a non-linear manner.
			We describe the formulation, model estimation, interpretability and examine the relative performance and econometric implications of our proposed model.
			We present an illustrative example of the model on a classic red/blue bus choice scenario example. 
			For a real-world application, we use a travel mode choice dataset to analyze the model characteristics compared to traditional neural networks and Logit formulations.
			Our findings show that our ResLogit approach significantly outperforms MLP models while providing similar interpretability as a Multinomial Logit model.
		\end{abstract}

	\end{frontmatter}
	\section{Introduction}
\label{sec:introduction}
Enhancing discrete choice models with neural nets and deep learning optimization algorithms is an active domain of research that has shown promising results \citep{sifringer2020enhancing,borysov2019generate,wong2020partite}.
In recent years, experimental use cases of deep learning methods in discrete choice modelling have been explored such as automatic utility discovery \citep{sifringer2020enhancing}, variational inference optimization \citep{bansal2019bayesian} and remapping explanatory variables into transferrable embeddings for travel behaviour modelling \citep{pereira2019rethinking}.
This paper provides a perspective of how a \textit{residual neural network} formulation accounts for unobserved choice heterogeneity in discrete choice models.
While the proposed model we have developed has its roots in the Mother Logit model, it is not a Random Utility Maximization (RUM) consistent model. Likewise, many non-RUM compatible models are used in discrete choice modelling that is still very useful \citep{hess2018revisiting}.

The increase in popularity of DNNs can be attributed to the general notion that these novel modelling strategies emulate behavioural actions and behaviour formation through neurological adaptations observed in the human brain \citep{bengio2015towards}.
This is referred to as `biological plausibility' in deep learning literature and is an efficient way of generating and representing decision models \citep{friston2007free}.
The similarity between behaviour theory and DNN has led to many interesting and useful applications in travel behaviour modelling and travel demand forecasting \citep{cantarella2005multilayer,lee2018comparison,wong2018discriminative,wang2019multitask}.
Intuitively, DNNs are made up of several layers of linear and non-linear operations, called activation functions, which enable the feasibility of estimation from noisy and complex data.

However, machine learning methods have their drawbacks.
Even though these methods are increasingly being studied in travel mode choice prediction ever since a decade ago \citep{karlaftis2011statistical}, their usefulness has been limited to prediction tasks, lacking the explainability of models.
Prediction accuracy as a comparison tool has been primarily used in early research in machine learning for travel behaviour modelling work and found to be that neural networks appear to lack consistency with economic principles \citep{hensher2000comparison}.
It is argued that DNN may not be suitable for econometric interpretation, and would lead to incorrect assumptions of the stochastic nature of decision-making behaviour.
More recent studies have compared the performance of discrete choice and machine learning in prediction.
Variable importance analysis has shown that, in most cases, DNNs outperform discrete choice models \citep{omrani2013prediction,hagenauer2017comparative,wang2018machine}.

It has been observed in machine learning models that increasing the number of layers beyond a specific limit would degrade the model due to overfitting, unreachable optimal solutions, and model identification problems \citep{glorot11deep,he2016residual}.
Even in cases showing DNNs producing more accurate predictions\footnote{Assuming discrete classification probabilities.} than discrete choice models, the structural formulations are not consistent across studies.
Another problem with DNNs, although less of immediate concern, is the inconsistency in meta-learning hyperparameter selection, data-leakage and illogically estimated parameters \citep{hillel2019machine}.
Although not covered in this study's scope, we can address these problems with regularization techniques such as Gradient Batch Normalization or Dropout, or adaptive gradient search such as Adam or AdaGrad \citep{kingma2014adam}.
Moreover, the applicability of machine learning algorithms has not yet been justified in behavioural modelling applications and economic analysis beyond ad-hoc decision tree\footnote{Note: Methods used to select the subset of features in a decision tree results in categories that are sometimes arbitrary. Tree splitting rules are ultimately ad-hoc heuristics. However, comparative selection methods may still be useful if used to inform analysts about which metrics to use in specific choice scenarios.} learning approaches, which are not robust and based on greedy heuristics that do not generalize well from training data \citep{witten2016data,brathwaite2017machine}.
Lastly, training and optimizing a multi-layered discrete choice model to capture variations in taste heterogeneity have not yet provided the expected benefits beyond few ``shallow'' layers \citep{wang2019multitask}.

This paper proposes a tractable method of incorporating a \textit{data-driven neural network architecture} into a random utility choice model. 
We seek to improve choice modelling methodologies by incorporating algorithms that work well for deep learning and can be used in choice modelling while performing post-estimation welfare analysis.
It extends the systematic utility function to include attributes of other alternatives in potentially non-linear ways to relax the independent and identically distributed (IID) assumptions.
The model structure is similar to the existing Mother Logit family of models that incorporate relaxation of the independence of irrelevant alternatives (IIA) property to account for correlation between the IID error terms and the observed explanatory variables \citep{mcfadden1977application,timmermans1992mother}.
Our strategy is inspired by the concept of Residual Neural Networks (\textit{ResNet}) in deep learning literature -- adding skip connections between layers allows gradient backpropagation across multiple layers to address the vanishing gradient problem \citep{bengio2015towards}.
Recent studies have shown that this strategy significantly improves the learning algorithm in deep neural network architecture with marginal or no loss in performance \citep{witten2016data,he2016residual}.
We show that we can easily adapt the ResNet approach for discrete choice models, and it has similarities to the Mother Logit utility formulation.
Our proposed methodology provides the utility function with a generic Deep Learning method of correcting for choice heterogeneity in the model using a residual function in the model formulation.
This allows one to leverage deep learning algorithms to estimate new choice models.
We define this new choice model structure as a \textit{ResLogit} model.

This paper aims to present a practical implementation of neural networks in choice modelling research that leverages the strengths of deep learning.
While this paper deals on the consistency with utility maximization methods, we acknowledge that there are other numerous methods in deep learning literature for optimization through regularization, hyperparameter search, meta-algorithms that are comparable in performance to our ResLogit implementation.
This study focuses on the methodological benefits of deep learning in discrete choice analysis.
Our work contributes to the use of deep learning methodology in travel behaviour modelling. It has since been highly relevant in today's context of data-driven modelling and use of Big Data for choice and behaviour modelling.
In summary, the main contributions of this work are:
\begin{itemize}
	\item We present the specification of the ResLogit model that uses a residual DNN error correction component in the choice utility in the form of a \textit{data-driven} choice model.
	\item We present the desirable effects of the ResLogit that enables parameter estimation tractability and interpretability due to the skipped connections between neural network layers and allows for econometric $\beta$-parameters to be estimated consistently.
	\item We analyze the role of residuals in econometric behaviour models and improve previous attempts to integrating deep learning methods in discrete choice applications.
\end{itemize}

This paper is organized as follows:
\Cref{sec:background} provides a primer of neural networks and an overview of discrete choice models.
\Cref{sec:specification} presents the specification of our proposed ResLogit model.
\Cref{sec:redblue} demonstrates our formulation on a classic red-bus, blue-bus example.
\Cref{sec:casestudies} evaluates the methodology on a real-world travel dataset and discusses the results.
Finally, \Cref{sec:conclusion} concludes our work and discusses future implications of incorporating deep learning techniques in discrete choice modelling.

	\section{Background}
\label{sec:background}
Logit models have traditionally been used to analyze relationships between observed behaviour and attributes associated with the choices and decision maker's characteristics \citep{ben1985discrete,ben1995discrete}.
This framework has proved successful for decades because of its parsimonious, tractable, and flexible model formulation for representing rational behaviour assumptions.
It assumes that the underlying decision processes are unknown from the observer, and decision-makers select their preferred choice by ranking all potential alternatives and choosing the alternative with the maximum utility through Random Utility Maximization (RUM) theory.
The modeller is assumed to have incomplete information about the decision-maker's behaviour, and the model will have to account for some uncertainty.

An important feature of the Logit model is the IIA property, which is an outcome of the assumption that the error terms of the alternatives in an MNL model are IID \citep{mcfadden1978modeling}.
When the error terms are correlated, strict IID assumption may lead to an incorrect forecast and model misspecification.
The Logit model imposes a random error term representing behavioural uncertainty and account for the lack of information presented to the analyst.
This random error term is assumed to be uncorrelated to the attributes of the alternatives.
Extensions to the Logit models such as Nested Logit and Mixed Logit have been developed to account for the error correlation when the assumption does not hold.

\subsection{Representation of non-linearity and cross-effects in choice utilities}
Model misspecification may arise when the error terms are correlated with non-chosen alternatives.
Various studies in discrete choice modelling have accounted for heterogeneity across choice alternatives and decision-makers by incorporating attributes of non-chosen alternatives known as \textit{cross-effects}.
The assumption is that the included additional function conditions for part of the error term correlate with the non-chosen alternatives.
There are several approaches to dealing with similarities and cross-effects between alternatives \citep{schuessler2007recent}:
\begin{itemize}
	\item Segmentation into nests or classes,
	\item Analyzing the variance-covariance structure, and
	\item Incorporating similarity factors into the deterministic part of the utility.
\end{itemize}

The first group consists of extensions to the MNL model such as the Nested Logit model to partially relax the IID assumption by segmenting alternatives into subsets. They are similar within each group (correlated) but independent between groups (non-correlated).
These models specify the correlation between alternatives by allowing attribute coefficients to vary between observations, class segments or individuals.
Although this model formulation works well with simple stated preference choice scenarios where the analyst can control the survey questions and options, cognitive bias formed during the behaviour learning process, e.g. anchoring effects, are not fully captured \citep{tversky1981framing}.
For instance, when a traveller makes a mode choice decision, there is a tendency to rely heavily on the information that they learn.
The learning process may also evolve, resulting in Spatio-temporal heterogeneity.

The second group consists of the Generalized Extreme Value (GEV) model family (e.g. Mixed Logit), and Probit models which allow for different (co-)variances among the error term in the utility function \citep{mcfadden1977application,daganzo1977multinomial}.
Multivariate distributed random error terms are introduced into the utility to capture potentially any correlation structure.
This assumption works well with simple behavioural models and allows for tractable estimation.
It does not necessarily reflect observed behaviour accurately with arbitrarily defined error distribution for more complex behavioural models.
However, we can also derive individual-specific estimates from the individual's conditional distribution based on their choices \citep{hensher2003mixed}.
Identification and computation of a large number of random distribution are still problematic in conventional discrete choice applications.
Recent research efforts have also focused on Mixed Logit estimation using optimization techniques primarily used in machine learning.
In particular, Bayesian variational inference optimization methods have shown to be promising \citep{bansal2020bayesian}.

The third group consists of models that include an explicit measure of similarity among alternatives in the utility function.
This group include hybrid choice models and the integrated choice and latent variable (ICLV) family of models.
Most notably, the Mother Logit model introduced by \citet{mcfadden1975independence} represents a generalization of the conventional MNL model, but not necessarily RUM consistent, by allowing for the existence of cross-effects and other substitutions (reference dependence, decoy, anchoring bias, regret, etc.) in the utility to relax the IID assumption \citep{timmermans1992mother}.

The Mother Logit formulation can approximate any discrete choice model in which the alternative's scale value is a function of all attributes of all choices \citep{timmermans1992mother}.
Other choice model development such as the Random Regret Minimization (RRM) model \citep{chorus2010new} which include terms from foregone alternatives, can be reformulated as a Mother Logit model \citep{mai2017similarities}.
The RRM model bases the assumption that one or more alternatives outperform the desired choice. 
This is translated into an anticipated regret function, and the analyst can formulate the non-linear utility as a function of attribute cross-effects of all the alternatives in the deterministic component and a random error term.
\citet{mai2017similarities} also presented a case of a Recursive Logit (RL) model based on the Mother Logit formulation.
\citet{mai2017similarities} formulated the RL model utility functions as a route choice problem, which computes the sum of the outgoing link utility and the expected maximum utility to the destination node, accounting for these cross-effects in the link utility functions.
When links overlap between different feasible route choice alternatives, the non-linear RL utility of a given route choice would include attributes from other route alternatives.

Cross-effect represents the utility correction measure of similarity or dissimilarity across all attributes of all alternatives \citep{timmermans1992mother}.
A negative cross-effect indicates that an IIA model overestimates the utility of the alternative due to correlated attributes and alternatives (e.g. the red/blue bus problem \citep{mcfadden1973conditional}).
Likewise, a positive cross-effect indicates that the utility is underestimated and a positive bias correction is required to account for the choice heterogeneity.
The Mother Logit formulation implies that the model violates RUM regularity conditions \citep{timmermans1992mother}.
Nevertheless, such model flexibility can accommodate behavioural anomalies incompatible with RUM based models \citep{hess2018revisiting}.

\subsection{Generalized approach to capture non-linearity and cross-effects in discrete choice models using DNNs}
\label{sec:gen_approach}
Passive data collected from sensors, devices and infrastructure that track decision making actions over time can reveal learning behaviour and trends of the decision-makers.
The general approach of representing decision-making uncertainties and learning processes as probabilistic error terms may be sufficient in obtaining satisfactory approximations. 
However, it is often difficult to identify the source of heterogeneity due to the complex interactions between influences from various attributes of non-chosen alternatives over a long period of interaction.
Furthermore, it provides no useful indication of selecting the error term mixing distribution or how many mixing distributions are required to reach an acceptable estimation of the decision-making behaviour \citep{mcfadden2000mixed}.

Combining DNNs and discrete choice modelling strengths have been explored in the past several years \citep{borysov2019generate,bansal2019bayesian,pereira2019rethinking,wong2020partite,badu2020composite}.
These new hybrid models are designed to capture learning behaviour and trends from large datasets, independent from the subjective bias induced from stated preference survey questionnaires.
The decision making learning algorithm is assumed to contain non-linear cross-effects, which results in complex error distributions and a non-linear utility function.
In practical choice modelling applications, the learning algorithm's process updates the model is unknown to the modeller.
Therefore it is said to be a `black-box' model \citep{breiman2001statistical}.
Non-linear activation functions in DNNs are assumed to represent taste variations and random heterogeneity in the choice model.
For instance, a non-compensatory decision protocol distribution is often used to generalize decision rules in discrete choice, rather than to define fixed assumptions about the error distribution \citep{vythoulkas2003modeling}.

Although neural networks have proved popular in recent years with their simple design and implementation, they rely on hyperparameter search or meta-learning process which cannot be intuitively interpreted from a micro-economic perspective.
Hyperparameters are the learning algorithm parameters that specify the learning procedure: $L_1$ and $L_2$ penalties, gradient step size, decay or initialization conditions.
In some situations, hyperparameter tuning\footnote{hyperparameter tuning refers to the specification of the \textit{learning algorithm}, not the model parameters, e.g. $\beta$ parameters.} can yield state-of-the-art performance.
\citet{lipton2018mythos} gave the hypothesis on lack of model interpretability by identifying that most machine learning-based systems may achieve high accuracy despite failing to explain where the source of the difference lie.
The MLP model is seen as a `black-box' model and will not be able to identify the beta parameters associated with the independent explanatory variables.
Model identifiability may be problematic as there can be multiple model specification defined by the same set of parameters.

\subsection{General formulation of a neural network model}
We explain the necessary notations and formulation of an MLP network, the \textit{ResNet} architecture, and how we can integrate the residual functions into a choice model, which follows a logically consistent extension of traditional MNL that relaxes the IIA property.

Each neuron in an MLP is a basic processing unit that performs a non-linear transform on the input \citep{lee2018comparison}.
The goal is to approximate some function $y=f^*(\B{V})$ with $y=f(\B{V};\theta)$, where the input $\B{V}$ is a linearized function of a vector of observed variables $\B{x}$ and a vector of estimated parameters $\BS{\beta}$, denoted as $\B{V} = f(\BS\beta,\B{x})$.
The function $f(\B{V};\theta)$ is a map of the linear components $\B{V}$ to a vector of discrete choice probabilities $y$.
$\theta$ is the neural network parameters that result in the best approximation of $f^*$.
During the training process, the model is estimated by a batched gradient descent algorithm given an objective function, i.e. maximum likelihood estimation\footnote{Batched gradient descent is most used in deep learning optimization. For most machine learning problems, the data size is too large for quasi-Netwon methods such as BFGS/L-BFGS algorithm to perform in \emph{comparable time}. Furthermore, computing in batches allows for parallelized computation on GPUs.}
\footnote{In general, the \emph{no free lunch theorem} in optimization states that no one solution works best for all problems}.
The MLP architecture can be represented mathematically as a series of chain functions:
\begin{align}
	\label{eq:mlp1}
	\begin{aligned}
	\B{h}^{(1)} &= f^{(1)}(\B{V}) \\
	\B{h}^{(2)} &= f^{(2)}(\B{h}^{(1)}) \\
	&\ldots \\
	\B{h}^{(M)} &= f^{(M)}(\B{h}^{(M-1)}) \\
	y &= softmax(\B{h}^{(M)})
	\end{aligned}
\end{align}

\noindent
where $f^{(1)}$, $f^{(2)}$, ..., $f^{(M)}$ are the activation functions of the DNN and $M$ gives the depth of the model.
For example, a 3-layer DNN results in the general form $f(\B{V})= f^{(3)}(f^{(2)}(f^{(1)}(\B{V})))$.
$\B{h}^{(1)},\B{h}^{(2)},...,\B{h}^{(M)}$ are the intermediary non-linear output of each $m^{th}$ activation function and the final layer is a \textit{softmax} function\footnote{This softmax function is equivalent to a conditional Logit in discrete choice problems.}, and the output results in a vector of discrete probabilities associated with each choice.
The choice of activation functions is loosely guided by neuroscience observations and `biological plausibility', which refers to the similarity between the behaviour theory and signal transmission in the nervous system \citep{goodfellow2016deep}.
The activation function can be linear or non-linear.
For example, using a sigmoid function: $f(\B{V})=(1+e^{-\B{V}})^{-1}$ results in a probabilistic output between 0 and 1.
In general, most DNN architectures suffers from non-identifiability due to the nature of the chain of non-linear activation functions -- the change in $\beta$ parameter associated with the explanatory variable cannot be mapped directly to the output probabilities.

The na\"{i}ve intuition is that the MLP can learn increasingly complex features by adding more layers, and each layer returns an ``improved'' approximation of $f^*$.
On the contrary, research has shown that the number of layers representing a perfect model does not follow an asymptotic limit. Still, it deteriorates as one increases the number of layers \citep{srivastava2015training,he2016residual}, contradicting the assumption that DNNs provides greater flexibility than conventional discrete choice models.
Observations in discrete choice literature affirm this technical limitation of using multiple deep layers to improve modelling accuracy \citep{alwosheel2018dataset,lee2018comparison}.

\subsection{Formulating the neural network as a dynamical system}
The \textit{ResNet} architecture was proposed by \citet{he2016residual} to overcome the limitations of the MLP model.
We can interpret the model as a discretization of a dynamical system that exploits the use of identity shortcuts to enable the flow of information across layers without causing model degradation from repeated non-linear transformations \citep{he2016residual}.
From an optimization perspective, the hypothesis is that it is easier to optimize ``a small change to the input rather than improving the entire layer of inputs at once'' \citep{he2016residual}.
This approach potentially provides an attractive possibility for modellers to retain the econometric variables and allows the neural network function to approximate the underlying error variance from a choice modelling perspective.
Furthermore, it has been proven that the \textit{ResNet} model architecture has no critical points other than the global minimum \citep{hardt2016identity}.

The \textit{ResNet} model $y=f(\B{V})$ is defined as the following series of functions:

\begin{align}
	\label{eq:resnet1}
	\begin{aligned}
	\B{h}^{(1)} &= f^{(1)}(\B{V}) + \B{V} \\
	\B{h}^{(2)} &= f^{(2)}(\B{h}^{(1)}) + \B{h}^{(1)} \\
	&\ldots \\
	\B{h}^{(M)} &= f^{(M)}(\B{h}^{(M-1)}) + \B{h}^{(M-1)} \\
	y &= softmax(\B{h}^{(M)})
	\end{aligned}
\end{align}

The \textit{ResNet} uses a skip connection mechanism (eq. \ref{eq:resnet1}) to the gradient to propagate through the layers, preventing the vanishing gradient problem \citep{he2016residual}.
The last line of \Cref{eq:resnet1} transforms the output of the final intermediate layer to a vector of probabilities using the \textit{softmax} function\footnote{For consistency with literature, we denote \textit{softmax} in the context of neural networks, and Logit in the context of discrete choice. However, both functions are mathematically equivalent}.
We can further generalize the \textit{ResNet} blocks as a series of recursive functions:

\begin{equation}
	\label{eq:resnet2}
	\B{h}^{(m)}=f^{(m)}(\B{h}^{(m-1)};\theta^{(m)}) + \B{h}^{(m-1)},\hspace{1em} \B{h}^{(0)} = \B{V}, \hspace{1em}\textrm{for}\hspace{1em}m=1,...,M
\end{equation}

where $\B{h}^{(0)}$ is the input after the initial linearization of the utility and $\B{h}^{(M)}$ is the output map before the \textit{softmax} function.
Approximating the parameters of the neural network $\theta^{(1)},\theta^{(2)},...,\theta^{(M)}$ is equivalent to solving for a series of linear discrete optimal control problem $U_m=f(V_m;\theta_m)+\varepsilon_m$.
We can also interpret $\B{h}^{(1)},...,\B{h}^{(M)}$ as a series of non-linear utility components that capture the cross-effects induced by similarity or overlap with the non-chosen alternatives.
If $f^{(m)}$ in eq. \ref{eq:resnet2} is large, it indicates the presence of cross-effects on the output probability.
If this value is close to zero for all $m$ (non-linear cross-effects not present), the model would collapse to a Logit model.
	\section{Specification of the ResLogit choice model}
\label{sec:specification}
\begin{figure}[!h]
	\centering
	\includegraphics[width=\textwidth]{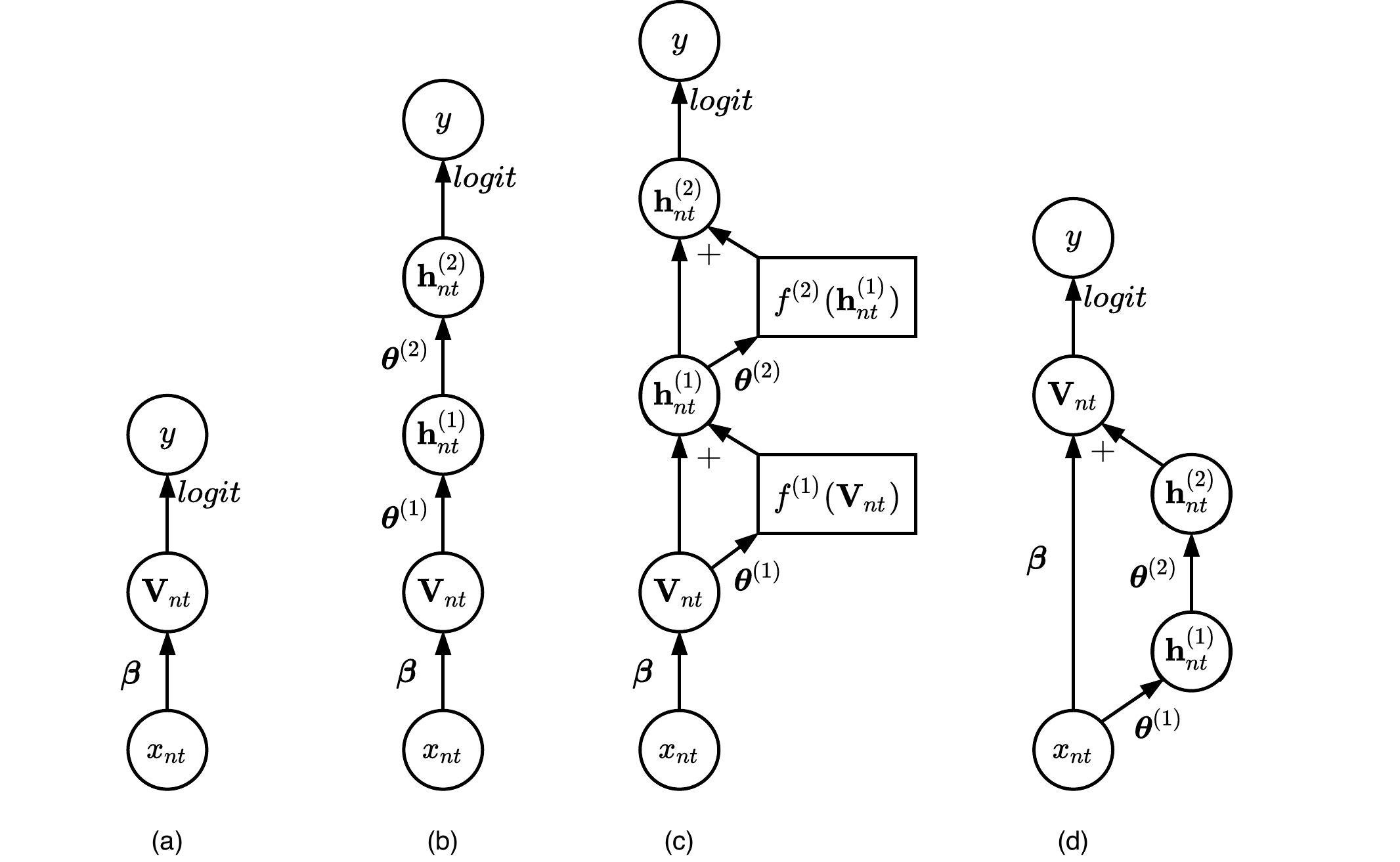}
	\caption{Simplified graphical model. (a) A Multinomial Logit model. (b) A MLP network with 2 hidden layers. (c) The proposed ResLogit model with 2 residual layers. Here we show the models expressed as symbolic operators that compute each step from the input $x_{nt}$ to the output probabilities $y$. The graph operator $+$ compute $h^{(m)}=h^{(m-1)}+f(h^{(m-1)})$. We omit the ASC variables for brevity. (d) Representation of the LMNL model used in \citet{sifringer2020enhancing}.}
	\label{fig:comp3}
\end{figure}

Our proposed ResLogit choice model improves discrete choice estimation by incorporating a neural network based on the recent \textit{ResNet} architecture.
\Cref{fig:comp3} shows a comparison between an MNL, MLP and the proposed ResLogit model as a simplified graphical model.
The general framework of our ResLogit architecture is that it is much more efficient to model the unobserved heterogeneity using a neural network rather than applying a neural network to the entire utility.
\citet{sifringer2020enhancing} applied a similar concept for a Learning MNL (LMNL) model, although using a fully connected neural network as a linear addition to the utility plus an unobserved error component.
This ad-hoc approach divided the explanatory variables into two groups, where one was used in the systematic linear utility and the other group in the neural network capturing the average effects.
In general, we specify the utility function as a sum of the deterministic component of observed characteristics and a neural network component that captures the unobserved heterogeneity in the choice process.
Our approach's advantage is that the skip allows for a greater chance of identifiability in the estimation of each layer of the neural network.
In contrast, the L-MNL model would still be vulnerable to the vanishing gradient problem.

A utility $U_{int}$ is defined by a deterministic component $V_{int}$ and a random error component $\varepsilon_{int}$:

\begin{equation}
	\label{eq:utility0}
	U_{int} = V_{int} + \varepsilon_{int}
\end{equation}

The deterministic component is a linear function of a vector of attributes $x_{nt}$ of a single alternative with a vector of estimated parameters $\boldsymbol\beta$.
The most general expression of the Logit model, the Mother Logit model, introduces a random variable $g_{int}$ in the utility that is a function of all attributes of all choices\footnote{Note to readers that the subscript $i$ refers to the index of the alternative in this section and the following sections. 
It does not refer to $g_{int}$ having only attributes from the $i^{th}$ alternative. We represent a function that depends solely on attributes from the alternative with an uppercase notation (e.g. $V$)}.
Note, in some cases, the random variable $g$ \emph{replaces} the deterministic part $V_{int}$ \citep{hess2018revisiting}.
Our ResLogit model's utility takes the general expression of the Mother Logit model as the output of the residual component.
The utility $U_{int}$ of individual $n$ selecting choice $i$ in a choice task $t$, from a choice set of $J$ alternatives with the residual component term is as follows:
\begin{equation}
	\label{eq:utility1}
	U_{int} = V_{int} + g_{int} + \varepsilon_{int}
\end{equation}

\noindent
The utility is a linear function of the systematic observed component $V_{int}$, the residual component $g_{int}$, and an extreme value distributed error term $\varepsilon_{int}$ representing the remaining unobserved errors not captured in the neural network.
$\B{V}_{nt}$ is a $J\times 1$ vector of utilities $v_{jnt}$ associated with each individual $n$ for choice task $t$:
\begin{equation}
	\B{V}_{nt} =
		\bordermatrix[{[]}]{
			 & \cr
			 & V_{1nt} \cr
			 & V_{1nt} \cr
			 & \vdots \cr
			 & V_{jnt}
		}_{J\times 1}
\end{equation}

\noindent
and $\B{g}_{nt}$ is a $J\times 1$ vector of residual components $g_{jnt}$ associated with the respective utility $j$ that contains all attributes from all alternatives:
\begin{equation}
	\B{g}_{nt} =
		\bordermatrix[{[]}]{
			 & \cr
			 & g_{1nt} \cr
			 & g_{1nt} \cr
			 & \vdots \cr
			 & g_{jnt}
		}_{J\times 1}
\end{equation}

\noindent
Eq. \ref{eq:utility1} would lead to the choice probability $y_i=f_i(\B{V},\B{g})$ for $i \in {1,...,J}$:
\begin{equation}
	\label{eq:condprob1}
	P(i) = y_i = \frac{\exp(V_{int} + g_{int})}{\sum_{j\in\{1,...,J\}}\exp(V_{jnt} + g_{jnt})}\hspace{1em}\forall i \in \{1,...,J\}
\end{equation}

\noindent where:

\begin{equation}
	\label{eq:utility_logsum}
	\B{g}_{nt} = -\sum_{m=1}^M \ln\left(1+\exp(\theta^{(m)}\B{h}_{nt}^{(m-1)})\right)
\end{equation}

\begin{equation}
	\B{h}_{nt}^{(0)}=\B{V}_{nt}
\end{equation}

\noindent
For any block $m$:
\begin{equation}
	\B{h}_{nt}^{(m)} = \B{h}_{nt}^{(m-1)} - \sum_{m'=1}^{m} \ln\left(1+\exp(\theta^{(m')}\B{h}_{nt}^{(m'-1)}\right), \hspace{1em}\textrm{for}\hspace{1em}m=1,...,M
\end{equation}

\noindent 
and $\theta^{(m)}$ is a $J\times J$ matrix of residual parameters:
\begin{equation}
	\theta^{(m)} =
		\bordermatrix[{[]}]{
			 & & & & \cr
			 & c_{11} & c_{12} & \dots & c_{1j'} \cr
			 & c_{11} & c_{22} & & \vdots \cr
			 & \vdots & & \ddots & \vdots \cr
			 & c_{j1} & \dots & \dots & c_{jj'}
		}_{J\times J}\hspace{1em}\textrm{for}\hspace{1em}m=1,...,M
\end{equation}

\noindent
where $c_{jj'}$ is the parameter matrix element for the $j^{th}$ row and $j'^{\textrm{ }th}$ column, and $\B{h}_{nt}^{(m)}$ is a $J\times 1$ vector of non-linear utility components for the $m^{th}$ residual layer:
\begin{equation}
	\B{h}_{nt}^{(m)} =
		\bordermatrix[{[]}]{
			 & \cr
			 & h_{1nt}^{(m)} \cr
			 & h_{2nt}^{(m)} \cr
			 & \vdots \cr
			 & h_{jnt}^{(m)}
		}_{J\times 1}\hspace{1em}\textrm{for}\hspace{1em}m=1,...,M
\end{equation}

The parameter matrices are defined such that the dimension of the residual output $\B{g}_{nt}$ matches the dimension of $\B{V}_{nt}$ for an element-wise additive operation.
We can have several intermediate neural network layers of varying sizes within each residual layer, which is one of the conveniences of the neural network architecture.
$\theta^{(m)}$ serves as the similarity or cross-effect factors to the utility function.
The chosen alternative's utility is increased or decreased by its degree of similarity with other non-chosen alternatives by this factor.
The MNL perspective corresponds to shifting the vector of utilities by $\B{g}_{nt}$.
If the cross-effect factors are zero, i.e. $\theta^{(m)}=0$ for all $m$, then the utility surplus is shifted by 0 and falls back to an MNL model.

Another observation is that the choice probability is conditional on the expectation of the output of the residual terms:
\begin{equation}
	\B{Q}_{nt}^{(m)}=\frac{1}{1+\exp(\theta^{(m)}\B{h}_{nt}^{(m-1)})},\hspace{1em}\textrm{s.t.}\hspace{1em}\B{Q}_{nt}^{(m)}\geq 0,\hspace{1em}\textrm{for}\hspace{1em}m=1,...,M
\end{equation}

\noindent
and if we assume that $\B{Q}_{nt}^{(m)}=\{Q_{jnt}^{(m)}\}$ for $j\in\{1,...,J\}$ is a vector of probabilities, we can rewrite the ResLogit formulation in \Cref{eq:condprob1} as a conditional choice probability:
\begin{equation}
	P(i) = y_i = \frac{\left(\prod_m Q_{int}^{(m)}\right) \exp(V_{int})}{\sum_{j\in\{1,...,J\}} \left(\prod_m Q_{jnt}^{(m)}\right) \exp(V_{jnt})}\hspace{1em},\forall i \in \{1,...,J\}
\end{equation}

The residual component (\Cref{eq:utility_logsum}) derives from entropy, or expected surplus function of the respective residual layers and the corresponding logsum term is the result of the log of the Logit probability denominator.
Behaviour modelling uses entropy to measure the variation or accessibility of a specific choice \citep{erlander2010cost}.
For example, \citet{mattsson2002probabilistic} characterized such formulation as maximization of the sum of the expected utility and a weighted entropy.
\citet{anas1983discrete} postulated that the entropy principle in choice models correspond to how much information-seeking behaviour is used to find the ``best'' utility specification.
\citet{fosgerau2017discrete} and \citet{matejka2015rational} also illustrated the affinity to generalized bounded rationality and the duality between discrete choice and rational inattention behaviour.
Consequently, information cost acts as a barrier between prior beliefs and the decision making actions, which results in choice heterogeneity.
An agent optimizes his or her desired outcome by minimizing this information cost \citep{matejka2015rational}.
Our ResLogit model aims to extend this concept by allowing for a data-driven surplus expression in the utility function (Presented in \Cref{eq:utility1}) to emulate the decision-makers' learning process.

\subsection{Depth of the neural network}
\label{subsec:depth}
Increasing the depth of the neural network increases the number of additive residual terms in the utility function.
The residual layers represent the underlying unobserved behaviour distribution that is not captured by the explanatory variables.
This mathematical formulation allows the model to reflect individual taste heterogeneities in the non-linear residual function.
Unlike a typical MLP model or the recently developed Learning-MNL model \citep{sifringer2020enhancing}, training a ResLogit model does not suffer from the vanishing gradient problem.
This eliminates the singularities caused by model non-identifiability.
This property's key implication on choice modelling is that we can operationalize the learning behaviour as a function in the utility while retaining the same econometric parameters in the structural equation.

\subsection{Estimation approach}
\label{subsec:estimation_approach}
The estimation procedure is a data-driven first-order stochastic gradient descent SGD learning algorithm, and we evaluate the performance on an out-of-sample validation set.
In data-driven optimization, we are maximizing a performance measure (e.g. out-of-sample performance) by indirectly maximizing a different surrogate objective function (e.g. maximizing log-likelihood of the training data).
We typically assume that the out-of-sample dataset is independent and identically distributed from the training dataset.
In contrast, pure optimization of discrete choice models directly maximizes the likelihood objective function, which is a goal of itself.
This method of estimating a large number of parameters has been proven efficient in machine learning.
In some cases, a surrogate objective function approach may result in a faster and better solution \citep{goodfellow2016deep}.
Other pre-conditioning methods or extensions can also be implemented into the surrogate objective function allowing it to reach multiple local optimum points and provide a regulating effect.
For example, these pre-conditioning includes adding momentum, adaptive learning rate methods or gradient noise normalization, see \citet{ruder2016overview} for an overview of such methods.
Another important difference is that the final convergence criteria are based on the performance measure, not the surrogate objective function within data-driven optimization.
This approach enables the algorithm to terminate when overfitting begins to occur (early-stopping criteria).
The estimation reaches convergence when the objective function no longer improves.

For this reason, a data-driven approach is more suitable in estimating our ResLogit model since a pure optimization approach will run into model non-identifiability issues due to a large number of estimated parameters.

\subsubsection{Objective function and parameter updates}
\label{subsec:objective_func}
The set of optimal parameters $\theta$ and $\BS{\beta}$ are estimated by maximizing the log-likelihood, where the log-likelihood is as follows:
\begin{equation}
	LL(\theta,\BS{\beta}) = \sum_{n=1}^N\ln P(i_n|\B{x}_n;\theta,\BS{\beta}).
\end{equation}

\noindent
The mini-batch SGD algorithm performs the following update rule on each iteration $t$:
\begin{align}
	\theta_{t+1} &= \theta_{t} - \eta_t \nabla_{\theta} \mathcal{J}_{\mathcal{B}}(\theta,\BS{\beta}),\\
	\BS{\beta}_{t+1} &= \BS{\beta}_{t} - \eta_t \nabla_{\BS{\beta}} \mathcal{J}_{\mathcal{B}}(\theta,\BS{\beta}),
\end{align}
\noindent where:
\begin{align}
	\nabla_{\theta}\mathcal{J}_{\mathcal{B}}(\theta,\BS{\beta}) = \frac{1}{K} \sum_{n'\in\mathcal{B}} \nabla_{\theta}LL_{n'}(\theta,\BS{\beta}),\\
	\nabla_{\BS{\beta}}\mathcal{J}_{\mathcal{B}}(\theta,\BS{\beta}) = \frac{1}{K} \sum_{n'\in\mathcal{B}} \nabla_{\BS{\beta}}LL_{n'}(\theta,\BS{\beta}),
\end{align}

\noindent
and $K$ is the batch size, $\mathcal{B}$ is a batch of observations sampled from $\B{x}_n$, $n'$ denotes the observation in the batch and $\eta_t$ is the learning rate.
We can regard $\nabla\mathcal{J}_{\mathcal{B}}(\theta,\BS{\beta})$ as a noisy estimate of the true gradient $\nabla LL(\theta,\BS{\beta})$.
We sample from the training set and adjust the $\BS{\beta}$ and $\theta$ parameters to reduce the training error, then we monitor the error in the validation by sampling from the validation dataset.
The goal of the optimization is to reduce the validation error while also reducing the difference between the training and validation error.
This can also be achieved by taking the model at the maximum log-likelihood of the validation dataset with an assumption that the estimation on the training dataset is asymptotic as the number of iterations on the samples $N\rightarrow \infty$.
The derivatives of the estimated parameters is computed using backpropagation \citep{goodfellow2016deep}.
Given the ResLogit formulation and taking the backpropagation from the output log-likelihood, the derivative of the log-likelihood with respect to $\BS{\beta}$ is:

\begin{equation}
	\frac{\partial LL}{\partial \BS{\beta}} = \frac{\partial LL}{\partial \B{V}}\frac{\partial \B{V}}{\partial \BS{\beta}} + \frac{\partial LL}{\partial \B{h}^{(m)}}\frac{\partial \B{h}^{(m)}}{\partial \BS{\beta}} + \frac{\partial LL}{\partial \B{h}^{(m-1)}}\frac{\partial \B{h}^{(m-1)}}{\partial \BS{\beta}} + ... \frac{\partial LL}{\partial \B{h}^{(1)}}\frac{\partial \B{h}^{(1)}}{\partial \BS{\beta}}
	\label{eq:bprop}
\end{equation}

The gradient formulation is shown in eq. \ref{eq:bprop} that by the nature of the residual connections, each derivative of the residual layers is independently computed.
This prevents the phenomena known as vanishing gradient.
If any of the gradients is computed to be zero, it does not affect the total backpropagated value and the $\BS{\beta}$ parameters can still be updated.
This allows the ResLogit to converge to an optimal MNL solution, even with non-identifiable residual layers.
In contrast, with a fully connected MLP model, the gradient formulation is a result of a chain rule:

\begin{equation}
	\label{eq:chainrule}
	\frac{\partial LL}{\partial \BS{\beta}} =
	\frac{\partial LL}{\partial \B{h}^{(m)}}\frac{\partial \B{h}^{(m)}}{\partial \B{h}^{(m-1)}}...\frac{\partial \B{h}^{(1)}}{\partial \B{V}}\frac{\partial \B{V}}{\partial \BS{\beta}}
\end{equation}

In eq. \ref{eq:chainrule}, if any of the intermediate derivatives are zero, then the total derivative is zero, and the model fails to learn and update $\BS{\beta}$, resulting in model non-identifiability.
The number of parameters used is relative to the number of alternatives in the choice set. Each element in the matrix corresponds to the cross-effects of other alternatives on the chosen alternative.
The diagonal elements in the matrix are the cross-effects with itself, i.e. a scale factor adjustment.
If this residual matrix is an identity matrix, that means that there are no cross-effects induced between alternatives (IIA holds), and the model collapses into a standard MNL model.
	\section{Red/Blue bus theoretical example}
\label{sec:redblue}
We show an example of how a simple nesting structure can be obtained using the ResLogit formulation in a hypothetical scenario.
Let us consider the red/blue bus problem.
The red/blue bus problem is a classic example of IIA property violation in choice models.
The problem arises in the assumption that the error terms for the red and blue bus options are independent, but they are correlated and share similar decision attributes in reality.
This means that the change in utility for a red bus will influence the change in utility of the blue bus.
To relax this assumption, choice modellers often use a Nested Logit model to relax the IIA assumption by adding a conditional probability term or logsum term.
The choice scenarios are summarized in \Cref{tab:choice_scenarios}.

\subsection{Scenario description}
\label{subsec:redblue_scenario}
In the first scenario (Scenario 1), assuming that we have a vector of 2 choices in a choice task $t$.
$\B{V}:\{ V_{car}, V_{bus}\}$, where each alternative has the same utility $V_{car}=1$, $V_{bus}=1$
Under strict IID assumptions, the probability of choosing either bus or car is, therefore, $P_{car}=P_{bus}=0.5$.

In the second scenario (Scenario 2), suppose that now we have a red bus $(V_{red\_bus})$ and blue bus $(V_{blue\_bus})$ option in place of $V_{bus}$, $\B{V}=\{V_{car}, V_{red\_bus}, V_{blue\_bus}\}$.
The utility of each alternative does not change, and all 3 alternatives have the same utility: $V_{car}=1$, $V_{red}=1$, $V_{blue}=1$.
Assuming the choice task is IID, the probabilities for the respective alternative should result in: $P_{car}=0.5$, $P_{red\_bus}=0.25$, and $P_{blue\_bus}=0.25$.
The probability of \textit{car} choice does not change when we add a new mode to the choice set.
However, the actual probabilities when estimated by an MNL model would result in: $P_{car}=0.33$, $P_{red\_bus}=0.33$, and $P_{blue\_bus}=0.33$, which does not seem plausible and violates IIA property conditions.

In the third scenario (Scenario 3), under our proposed ResLogit model, the correlation between the red and blue bus is corrected by a residual vector $\B{g}$, with residual parameter matrix $\theta^{(1)}$.
Using a 1-layer ResLogit model and a residual vector function defined by $\B{g}=-\ln(1+\exp(\theta^{(1)}\B{V}))$, we simulate a choice scenario with alternatives \textit{car}, \textit{red bus}, \textit{blue bus}.

We assume at a value of $1$ represents a positive cross-effect and a $-1$ value denotes a negative cross-effect and $0$ value represents no cross-effects (IIA property holds).
The negative value of cross-effects between the car and bus option may suggest that the alternatives are competing options (e.g. buses and cars sharing the same road segment).
we assign a value of $\{1\}$ to elements $c^{(1)}_{32}$ and $c^{(1)}_{23}$ and a value of $\{-1\}$ to elements $c^{(1)}_{12}$, $c^{(1)}_{21}$, $c^{(1)}_{13}$ and $c^{(1)}_{31}$:

\begin{equation}
	\label{eq:matrix01}
	\theta^{(1)} =
		\bordermatrix[{[]}]{
			& & & \cr
			& c_{11} & c_{12} & c_{13} \cr
			& c_{21} & c_{22} & c_{23} \cr
			& c_{31} & c_{32} & c_{33}
		} =
		\bordermatrix[{[]}]{
			 & & & \cr
			 & 0 & -1 & -1 \cr
			 & -1 & 0 & 1 \cr
			 & -1 & 1 & 0
		}.
\end{equation}

Given a $3\times 1$ vector of utilities $\B{V}=\begin{bmatrix}1&1&1\end{bmatrix}^{\top}$, the residual vector $\B{g}$ is:

\begin{align}
	\B{g} &= -\ln\left( 1 + \exp(\theta^{(1)}\B{V}) \right), \\
		 &= -\ln\left( 1 + \exp\Big(
		\begin{bmatrix}
			0 & -1 & -1 \\
			-1 & 0 & 1 \\
			-1 & 1 & 0
		\end{bmatrix}\cdot
		\begin{bmatrix}
			1 \\ 1 \\ 1
		\end{bmatrix}
		\Big) \right), \\
		&=
	\begin{bmatrix}
		-0.127 \\
		-0.693 \\
		-0.693
	\end{bmatrix},
\end{align}

\noindent
giving the choice probabilities as:

\begin{align}
	\begin{split}
		\label{eq:redblue_result}
		P(i) = \frac{\exp(V_i + g_i)}{\sum_{j\in C} \exp(V_j + g_j)} \hspace{1em} \textrm{for} \hspace{1em} i \in \textrm{\textit{car, red bus, blue bus}}\\
		P(\textrm{\textit{car}}) = 0.468; \hspace{1em} P(\textrm{\textit{red bus}}) = 0.265; \hspace{1em}P(\textrm{\textit{blue bus}}) = 0.265; \hspace{1em}
	\end{split}
\end{align}

The probabilities in \Cref{eq:redblue_result} show that with an addition of the residual matrix to account for the cross-effects, we have moved the choice probabilities of the car and red/blue bus options toward the true IIA conditions without changing the underlying utilities.

Now, if we assume no cross-effects between the car and bus alternatives (both car and buses are not sharing the same road segment), we update \Cref{eq:matrix01} with values of $\{0\}$ for parameters $c^{(1)}_{12}$, $c^{(1)}_{21}$, $c^{(1)}_{13}$ and $c^{(1)}_{31}$:

\begin{equation}
	\label{eq:matrix02}
	\theta^{(1)} =
		\bordermatrix[{[]}]{
			& & & \cr
			& c_{11} & c_{12} & c_{13} \cr
			& c_{21} & c_{22} & c_{23} \cr
			& c_{31} & c_{32} & c_{33}
		} =
		\bordermatrix[{[]}]{
			 & & & \cr
			 & 0 & 0 & 0 \cr
			 & 0 & 0 & 1 \cr
			 & 0 & 1 & 0
		}.
\end{equation}

\noindent
The resulting residual vector would be:

\begin{equation}
	\B{g} =\begin{bmatrix}
		-0.693 \\
		-1.313 \\
		-1.313
	\end{bmatrix},
\end{equation}

\noindent
giving the choice probabilities as:

\begin{align}
	\begin{split}
		\label{eq:redblue_result2}
		P(i) = \frac{\exp(V_i + g_i)}{\sum_{j\in C} \exp(V_j + g_j)}, \hspace{1em} \textrm{for} \hspace{1em} i \in \textrm{\textit{car, red bus, blue bus}}\\
		P(\textrm{\textit{car}}) = 0.482; \hspace{1em} P(\textrm{\textit{red bus}}) = 0.259; \hspace{1em}P(\textrm{\textit{blue bus}}) = 0.259; \hspace{1em}
	\end{split}
\end{align}

\begin{table}[!t]
	\caption{Illustration of red/blue bus choice scenario showing the effect of residual correction factors of a 1-layer model.}
	\label{tab:choice_scenarios}
	\centering
	\begin{tabu}{X[1] X[1,r] X[1,r] X[1.5,r] X[1,r]}
		\toprule
		Choice & $V_i$ & $g_i$ & $\exp(V_i + g_i)$ & $P(i)$ \\
		\midrule
		\multicolumn{5}{l}{Scenario 1} \\
		car & 1 & - & 2.718 & 0.5 \\
		bus & 1 & - & 2.718 & 0.5 \\
		\midrule
		\multicolumn{5}{l}{Scenario 2} \\
		car & 1 & - & 2.718 & 0.33 \\
		red bus & 1 & - & 2.718 & 0.33 \\
		blue bus & 1 & - & 2.718 & 0.33 \\
		\midrule
		\multicolumn{5}{l}{Scenario 3 (competing car/bus)} \\
		car & 1 & -0.127 & 2.394 & 0.468 \\
		red bus & 1 & -0.693 & 1.359 & 0.265 \\
		blue bus & 1 & -0.693 & 1.359 & 0.265 \\\\[-1em]
		\multicolumn{5}{l}{Scenario 3 (non-competing car/bus)} \\
		car & 1 & -0.693 & 1.359 & 0.482 \\
		red bus & 1 & -1.313 & 0.731 & 0.259 \\
		blue bus & 1 & -1.313 & 0.731 & 0.259 \\
		\bottomrule
	\end{tabu}
\end{table}

In principle, the nests between the car and the bus options are not pre-specified \textit{a priori} by the modeller.
The parameter matrix is estimated from data and defines the nesting structure or error term correlation of the choice alternatives.
The first observation of the hypothetical example shown above is that with a logical assumption of positive ($c_{bus,bus}=1$) cross-effect residual parameter between the two bus alternatives and zero ($c_{bus,car}=0$) cross-effect residual parameter between the car and bus alternatives would result in a nesting structure which reflects the relaxed IID assumption probabilities.
The second observation stems from the correlations between error terms of competing alternatives.
If the residual parameters are negative, it accounts for competing alternatives (e.g. buses and cars share the same road segment from the origin to destination), resulting in a slightly different outcome than a non-compete scenario.

	\section{Case study}
\label{sec:casestudies}
This study evaluates our proposed ResLogit model's effects and performance in three criteria: model depth, model degradation, and model predictive performance compared to an MLP neural network.
We also evaluate the residual effects on econometric parameters by comparing the beta and standard error values with a baseline MNL model without the residual layers.

We evaluate the ResLogit model's performance using individualized characteristics and attributes in a revealed preference (RP) travel survey dataset using out-of-sample accuracy at the minimum validation loss point on the validation curve.
We computed the accuracy using a 30\% hold-out validation set from our dataset.
We compared the model degradation effects between our ResLogit and a vanilla MLP model with identical model hyperparameters to address the adverse impact of model degradation from increasing layers.
We showed the effects of increasing layers in the ResLogit model and the MLP model on estimation accuracy and model identifiability.

\subsection{Data and model description}
\label{subsec:mtltrajet}
We used the 2016 \textit{Mtl Trajet} RP dataset collected from the user's smartphone data on a mobile application \citep{yazdizadeh2017generic}.
A list of explanatory variables and the choice set used for this mode choice prediction analysis are shown in \Cref{tab:mtltrajet}.
The respondents' travel diary includes mode choice, activity choice, trip attributes (e.g. trip length, start/end time, location) and GPS trajectories.
The travel survey was conducted over four months, from September to December 2016.
In total, there were 60,365 unique trips made during the period.
To evaluate out-of-sample performance, we divide the dataset into two sets using a 70:30 training/validation split ($N_{training}=42,256$ samples, $N_{validation}=18,109$ samples).
We developed the model estimation algorithm using open-source deep learning libraries in Python. 
The code for our experiments is available on our Github page\footnote{\url{https://github.com/LiTrans/reslogit-example}}.

\begin{table}[!t]
    \centering
    \caption{Descriptive variables of the dataset.}
    \label{tab:mtltrajet}
    \begin{tabu}{X[0.4,l] X[0.8,l] X[0.45,l] X[0.15,r] X[0.25,r]}
    \toprule
    variable & description & type & mean & std dev \\
    \midrule
    weekend & trip on weekend & dummy variable & 0.205 & 0.001 \\
    hour\_8\_10 & trip between 8am to 10am & dummy variable & 0.163 & 0.0015 \\
    hour\_11\_13 & trip between 11am to 1pm & dummy variable & 0.147 & 0.001 \\
    hour\_14\_16 & trip between 2pm to 4pm & dummy variable & 0.209 & 0.002 \\
    hour\_17\_19 & trip between 5pm to 7pm & dummy variable & 0.249 & 0.002 \\
    hour\_20\_22 & trip between 8pm to 10pm & dummy variable & 0.095 & 0.001 \\
    hour\_23\_1 & trip between 11pm to 1am & dummy variable & 0.03 & 6e-4 \\
    hour\_2\_4 & trip between 2am to 4am & dummy variable & 0.006 & 3e-4 \\
    hour\_5\_7 & trip between 5am to 7am & dummy variable & 0.101 & 0.005 \\
    num\_coord & number of trajectory links & continuous & 109.8 & 131.23 \\
    trip\_dist & trip distance (km) & continuous & 8.366 & 10.42 \\
    trip\_duration & trip duration (min) & continuous & 24.04 & 20.97 \\
    trip\_avgspeed & trip average speed (km/h) & continuous & 22.503 & 18.815 \\[1mm]
    activity & trip activity type:\{1: education, 2: health, 3: leisure, 4: meal, 5: errands, 6: shopping 7: home, 8: work, 9: meeting\} & categorical \\[1mm]
    \midrule
    choice alternatives & \multicolumn{4}{l}{1: Auto, 2: Bike, 3: Public Transit, 4: Walk, 5:Auto+Transit, } \\
           & \multicolumn{4}{l}{6: Other mode, 7: Other combination} \\
    \bottomrule
    \end{tabu}
    \end{table}

We iterated over the experiment by varying the depth of the ResLogit and MLP neural network using 
2, 4, 8 and 16 hidden layers $(M=\{2,4,8,16\})$.
Note that our study only shows a relative comparison between the models with a similar number of layers and neural network hyperparameters.
The objective of this experiment is to show the effectiveness of the ResLogit approach as a way of incorporating deep learning methods into discrete choice models over a conventional MLP neural network.

This experiment considers three specific objectives:
\begin{enumerate}
    \item Effects of the number of residual layers on the model $\beta$ parameters.
    \item Model validation accuracy and maximum log-likelihood estimation comparison.
    \item Comparison of estimated $\beta$ parameters between the ResLogit model and MNL model.
\end{enumerate}

The model estimation process begins with a baseline MNL estimation. 
Next, the MLP models were estimated (4 models, one each for 2, 4, 8 and 16 hidden layers), and labelled as MLP-2, MLP-4, MLP-8 and MLP-16, respectively.
We performed the same training process on the ResLogit models (RL-2, RL-4, RL-8, RL-16).
For the learning algorithm, we used the mini-batch SGD learning algorithm with a mini-batch size of 64 (i.e. gradient is computed over a sample of 64 observations from the training dataset) to train our models.
For the learning algorithm, we applied an RMSprop optimization step \citep{goodfellow2016deep}.
The ResLogit model residual parameters are initialized using an identity matrix.
Once the models have been trained, we take the best-specified model at the minimum validation loss point and compute the prediction accuracy using the validation dataset's model parameter values. 

\subsection{Analysis of model results}
\Cref{fig:curves_resnet_vs_mlp} and \Cref{fig:loss_curves_resnet_vs_mlp} reports the validation results of the MNL and ResLogit models with a baseline comparison to a MNL model (red line).
A condensed version of the estimated $\beta$ parameters of the MNL and ResLogit models are presented in \Cref{tab:mnl_residual}, which we showed the comparison between our best estimated ResLogit structure (RL-16) and the MNL model.
\Cref{fig:resmat} shows the parameters of the first four residual layers.

\subsubsection{Performance measure on out-of-sample data}
\begin{figure}[!h]
    \centering
    \includegraphics[width=\textwidth]{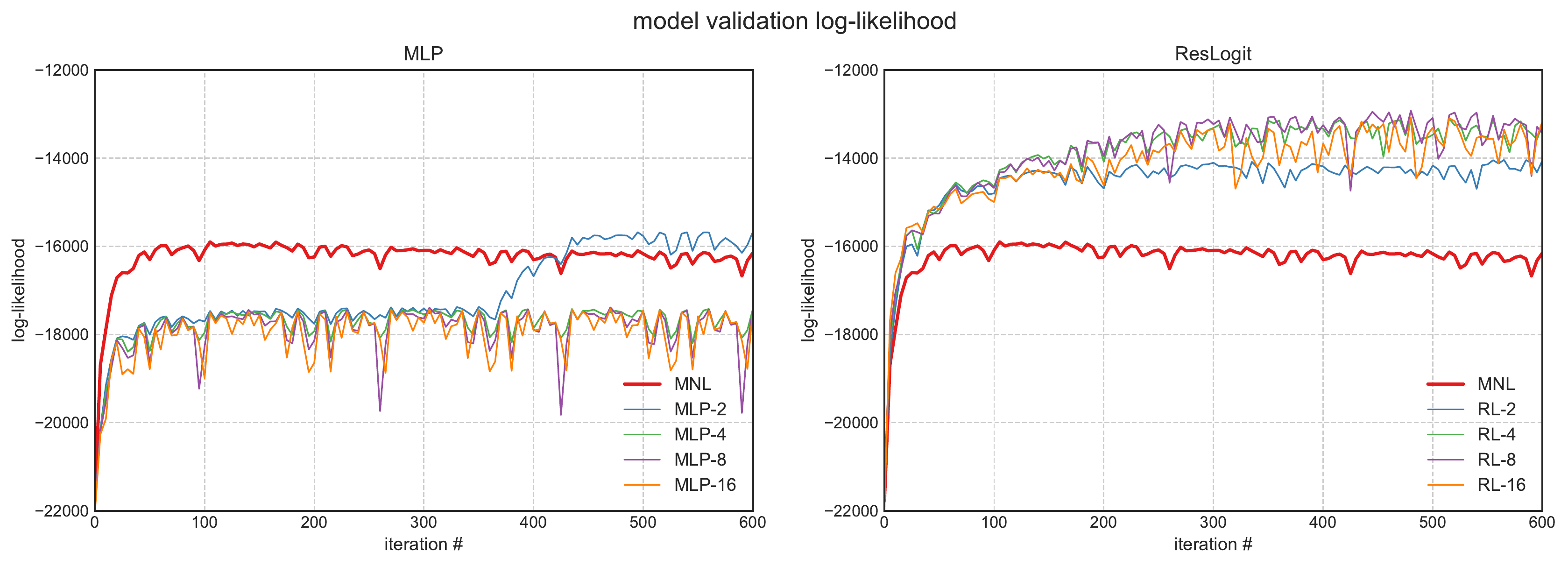}
    \caption{Validation log-likelihood results of the model estimation.}\label{fig:curves_resnet_vs_mlp}
\end{figure}

\Cref{fig:curves_resnet_vs_mlp} shows the validation curves of the model log-likelihood.
The x-axis represents the iteration step, and the y-axis reports the log-likelihood.
The MNL curve indicates the baseline performance where no augmentation to the utility or model.
The plot on the left shows the comparison between the MNL and MLP models.
This result indicates that the MLP model performs \textit{worse} than the MNL model. 
The only change between the MLP and ResLogit experiments is the model structure. Therefore the improvement is most likely only attributed to the change in model structure, and not other hyperparameters\footnote{It is also plausible that an MLP will do better or equivalent to a Logit model and sometimes an MLP can perform worse than a Logit model (on this particular class of problem, for example). This can be explained by the "No Free Lunch" theorem \citep{kawaguchi2017generalization}: "If an algorithm performs well on a certain class of problems, then it necessarily pays for that with degraded performance on the set of all remaining problems." \citep[Theorem 1]{wolpert1997no}.}.
MLP-2 also took twice as long to reach the maximum log-likelihood (400 vs 200 iterations on the MNL model). 
The MLP models (MLP-4, MLP-8 and MLP-16) produced significantly noisier output in the backpropagation step in SGD, which causes the ``spikes'' seen on the left plot.
There were also identifiability problems with MLP-4, MLP-8 and MLP-16 models. 
Since the MLP-4, MLP-8 and MLP-16 models were misspecified, they could not reach the same performance log-likelihood compared to the MNL models.
This result showed that adding neural network layers does not guarantee better performance and a simple MNL could potentially outperform a DNN, which is in line with our initial hypothesis.

We observed that as we increase the depth of the ResLogit models (\Cref{fig:curves_resnet_vs_mlp}, right), the log-likelihood remains consistent and outperforms the baseline MNL. 
Although we are using the same number of parameters and the same learning algorithm, the ResLogit method generated correctly specified models while the MLP models were misspecified. 
Model specification test is handled by out-of-sample validation analysis and econometric interpretation of beta parameters (explained in the following sections).
We note that we did not implement any other forms of regularization for experiment consistency, e.g. $L_1$, $L_2$ regularizer or Dropout techniques.
An alternative approach to model selection for more complex data where there are many unknown variables is to use a statistical measure such as the Akaike Information Criterion (AIC).
The AIC statistic calculated for the MNL, MLP-16 and RL-16 models is 32566, 34902 and 28086 respectively. 

\begin{figure}[!t]
    \centering
    \includegraphics[width=\textwidth]{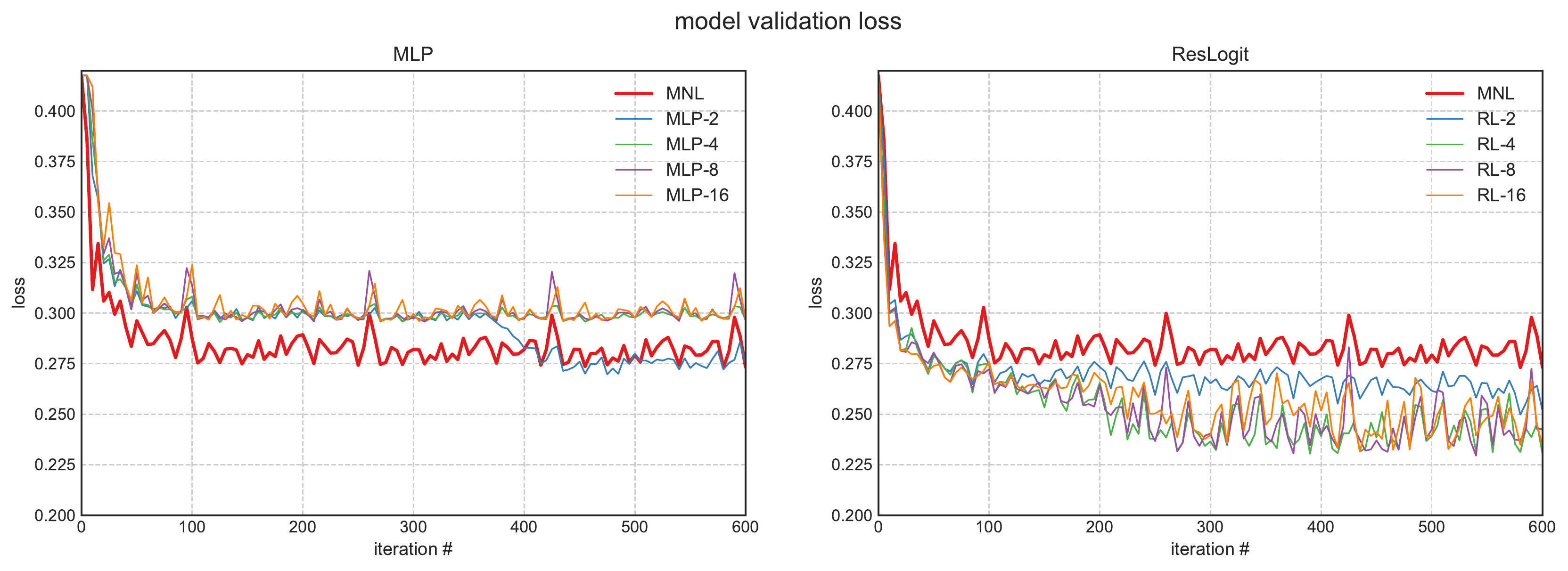}
    \caption{Validation loss comparison between the MLP models and the ResLogit models.}\label{fig:loss_curves_resnet_vs_mlp}
\end{figure}

\Cref{fig:loss_curves_resnet_vs_mlp} shows the validation error curves for both models. 
The error is defined as (1 - \textit{mean prediction accuracy}) where the \textit{mean prediction accuracy} is:

\begin{equation}\mathcal{L}_n(i,i^*)=
\begin{cases}
  1 \hspace{1em} i=i^*\\    
  0 \hspace{1em} i\neq i^*    
\end{cases}i,i^* \in \mathcal{D}_{validation}
\end{equation}
\begin{equation}
    \textrm{\textit{mean prediction accuracy}} = \frac{1}{N_{validation}}\sum_{n=1}^N \mathcal{L}_n(i,i^*)
\end{equation}

\noindent
where $i$ is the actual choice, $i^*$ is the predicted choice, $\mathcal{L}_n(i,i^*)$ is the 0-1 loss function and $\mathcal{D}_{validation}$ is the validation dataset.

The stability of convergence shows no strong overfitting bias during the estimation process.
On the MLP curves on the left plot, the model with the smallest error is the one with the least number of hidden layers but only after iteration 400, with the MNL model coming in as the second-lowest error.
We can see that the error reaches a saturation point around 0.3 for MLP-2 with a negligible decrease at MLP-4 to MLP-16.
This makes sense because the non-linear structure of the multi-layered neural network will be susceptible to the vanishing gradient problem observed in this figure. 
The results are more profound when we compare the MLP with the ResLogit model (\Cref{fig:loss_curves_resnet_vs_mlp}, right). 
In the MLP plot, we observe that the learning gets trapped in a locally optimal point.
The difference is minimal with two layers, which we expected, but a more pronounced difference between the MLP and ResLogit model when the number of layers increases.
On the right plot of \Cref{fig:loss_curves_resnet_vs_mlp} the loss gets progressively smaller as we increase the number of residual layers, which is consistent and follows a logical pattern.
Even with an RL-2, the error drops significantly faster, and the model achieved lower error than the MNL model as soon as the estimation starts.
This means that neural networks are best suited to capture the error distribution rather than using it as a transformative operator on the explanatory variables.

\subsection{Model coefficient estimates}
\Cref{tab:mnl_residual} presents the coefficient estimates, standard errors and robust standard errors for the observed explanatory variables for the MNL and RL-16 model.
The parameter estimates indicate the individuals' exhibited preferences for each attribute for each alternative.
The results show that individuals reacted towards a stronger preference for transit when the trip time is longer in the ResLogit model, relative to the MNL model.
Individuals also prefer a longer route for transit compared to auto according to the ResLogit model. In contrast, the MNL estimates show that individuals prefer a longer route when taking auto over transit.
There are specific indicators which are captured in ResLogit and not in the MNL models.
For instance, on weekends, people in Montreal use their car more to do shopping, recreation, visit their parents in the suburbs, go to cottage, etc. Therefore, ResLogit is giving us a positive sign for car over the weekend compared to other modes.
Another example is that during morning rush hour (8-10), people commute and there is a higher chance that they take auto+transit (due to the availability of a large amount of parking at stations) to reach their office. 
This fact is captured only by ResLogit.

Standard errors can be calculated through the Fisher Information Matrix, requiring only the Hessian of the log-likelihood which assumes a correctly specified model.
Additionally, the correct specification assumption can be relaxed by computing the robust sandwich estimator.
We calculate the standard errors as a function of the negative inverse of the Hessian matrix $\mathcal{H}$, which gives the variance-covariance matrix of $\beta$, assuming those estimates are normally distributed. 
This value gives the Cramer-Rao bound:
\begin{equation}
    \hat{\sum}_\beta^{CR} = - \hat{\mathcal{H}}^{-1}
\end{equation}

The Hessian matrix is the second-order derivative of the log-likelihood with respect to the model parameters.
Then, taking the diagonal of the square root of that variance-covariance matrix normalized by the size of the dataset, we obtain the standard errors.
The robust standard error $\hat{\sum_\beta^{Rob.}}$ is calculated by:
\begin{equation}
    \hat\sum_\beta^{Rob.} = (- \hat{\mathcal{H}}^{-1})\hat{B}(- \hat{\mathcal{H}}^{-1})
\end{equation}
where $\hat{B} = \sum_{n=1}^N(\frac{\partial LL_n}{\partial\B{\beta}})(\frac{\partial LL_n}{\partial\B{\beta}})^{\top}$
In terms of the coefficient significance value, the ResLogit parameters have more parameters with a nominal p-value < 0.05 compared to the MNL model.

The standard error and robust standard error estimates show that the ResLogit estimates are more reliable than the MNL model.
For the extreme cases, the parameter estimates for trip distance for walking showed the smallest value compared to other modes for both models as expected, indicating that the results are consistent.
The robust standard errors also show that some parameters are not significant, for instance, \textit{meeting activity-bike} has a high standard error when accounting for model misspecification. This is logical as travelling by bicycle is not usually common.
The estimates for \textit{hour (20-22)-bike} also indicate that this parameter is not a significant parameter, we can say that the hours between 8 and 10 pm does not impact the preference of \textit{bike} mode.

We caution the readers that we can give no general guarantees to the precision of the standard errors or the asymptotic behaviour of the model fit for heavily biased models \citep{goeman2018l1}, such as L1 or L2 regularization used in neural networks and other machine learning methods. 
Our ResLogit formulation reduces this bias in the model through the addition of residual layers to account for the systematic errors.
Therefore, the robust standard errors that we report are reliable, but only provide an approximation of model specification correctness and the variance of the estimates.

\subsection{Analyzing cross-effects from the residual matrices}
\begin{figure}[!t]
    \centering
    \includegraphics[width=\textwidth]{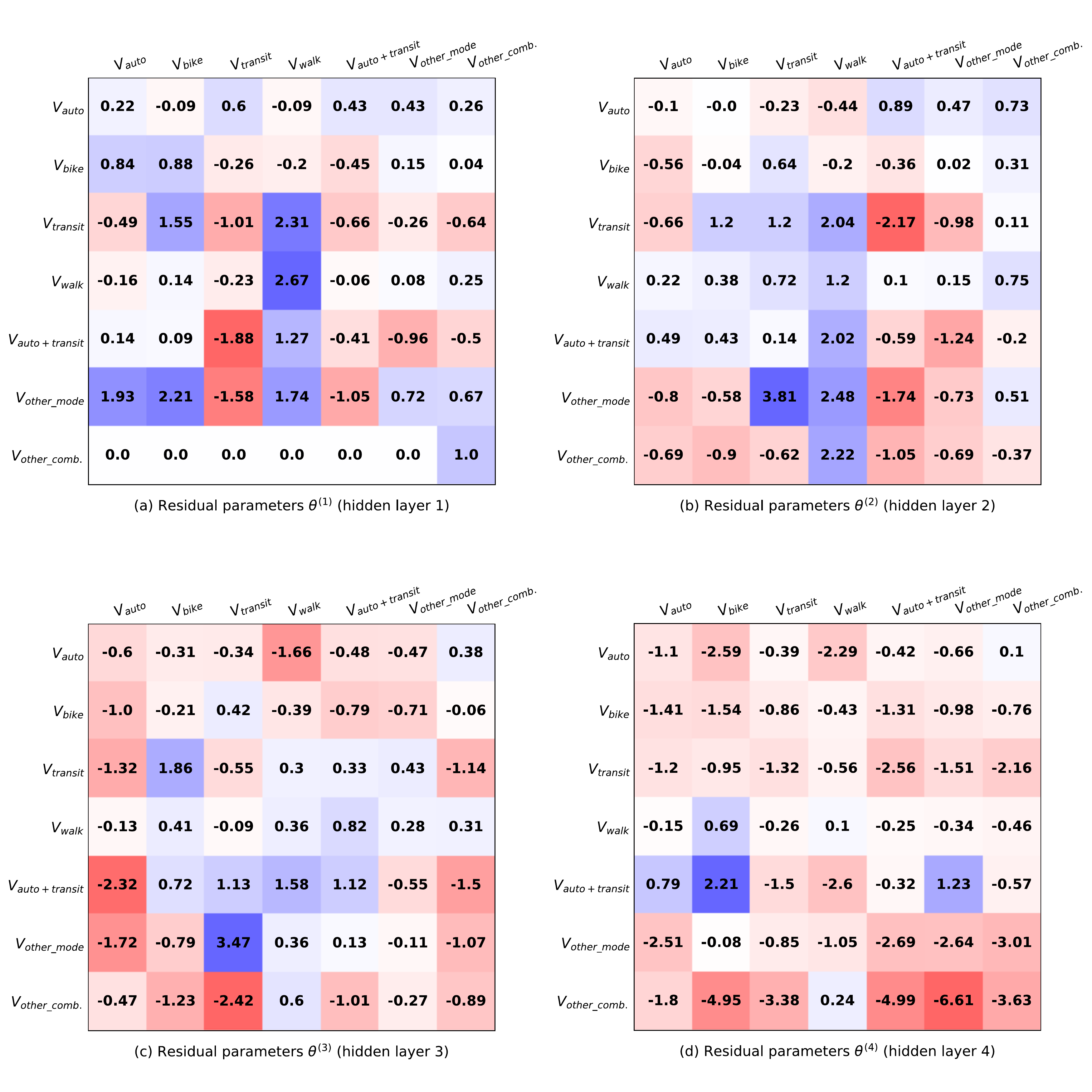}
    \caption{First 4 layers of weight matrices from the ResLogit model.}
    \label{fig:resmat}
\end{figure}

The cross-effects of non-chosen alternatives are reflected in \Cref{fig:resmat}.
The figure shows the parameters of the first four residual layers of RL-16.
The matrices' values correspond to the level of dependency between the utility of one alternative with the utility function of the second alternative, and vice versa.
As explained in \Cref{sec:redblue}, this matrix defines the underlying error term correlations between the choice alternatives. 
For example, the positive value of transit-bike in \Cref{fig:resmat} (a) is 1.55. This means that the attributes of transit mode positively influence individuals who choose bike mode, increasing the utility of transit influences the increase in mode share for the bike.
However, the reverse may not be identical.
The value for bike-transit in \Cref{fig:resmat} (a) is -0.26, indicating that increasing the utility of bike (e.g. more bike infrastructure), decreases the mode share for transit.
We may relate this observation to the shared infrastructure between auto and bike.
The non-zero values indicate the existence of non-linear cross-effects in the stated choices.
This analysis provides an estimate of the cross-effect influence between modes of travel.
Nonetheless, this experiment has shown how the ResLogit formulation uses the residual function to enhance model performance.

\begin{table}[!t]
\centering
\caption{Comparison of a subset of parameter estimates between MNL and ResLogit model.}
\label{tab:mnl_residual}
\begin{tabu}{X[2,l]X[1.5,l]X[1,r]X[1,r]X[1,r]X[1,r] X[1,r]X[1,r]}
\toprule
&& \multicolumn{3}{l}{MNL} & \multicolumn{3}{l}{ResLogit (16-layer)} \\
Parameter ($\beta_{mj}$)& Choice  &  parameter &  std. err. &  rob. std. err. &  parameter &  std. err. &  rob. std. err. \\
\midrule
weekend 		& auto 			&     -0.057$^*$ &      0.036 &           0.386 &      0.045$^*$ &      0.006 &           1.157 \\
				& bike 			&     -0.990$^*$ &      0.081 &           7.335 &     -0.448$^*$ &      0.063 &           7.566 \\
				& transit 		&     -0.751$^*$ &      0.042 &           1.569 &     -0.090$^*$ &      0.007 &           0.089 \\
hour\_8\_10 	& walk 			&     -0.841$^*$ &      0.070 &           7.986 &     -1.459 &      0.013 &           0.063 \\
				& auto+transit 	&     -2.273$^*$ &      0.121 &          15.005 &      1.162 &      0.032 &           0.230 \\
hour\_11\_13 	& bike 			&     -0.854$^*$ &      0.073 &          47.886 &     -1.210$^*$ &      0.071 &          15.565 \\
				& auto+transit 	&     -2.540$^*$ &      0.217 &          48.866 &      1.618 &      0.039 &           0.359 \\
hour\_17\_19 	& auto 			&      0.058$^*$ &      0.029 &           0.186 &     -0.586 &      0.004 &           0.001 \\
hour\_20\_22 	& bike 			&     -1.271$^*$ &      0.092 &          16.937 &     -0.943$^*$ &      0.085 &          15.009 \\
trip\_dist 		& auto 			&      0.354 &      0.007 &           0.002 &     -0.113 &      0.001 &           0.000 \\
				& transit 		&      0.297 &      0.008 &           0.002 &      0.817 &      0.001 &           0.000 \\
				& walk 			&     -2.197 &      0.028 &           0.387 &     -0.257 &      0.004 &           0.001 \\
trip\_time 		& auto 			&     -0.627 &      0.005 &           0.000 &     -0.397 &      0.001 &           0.000 \\
				& transit 		&      0.870 &      0.005 &           0.000 &      0.303 &      0.001 &           0.000 \\
				& walk 			&      0.863 &      0.009 &           0.007 &     -0.752 &      0.002 &           0.000 \\
trip\_aspeed 	& auto 			&      0.988 &      0.005 &           0.001 &     -0.024 &      0.001 &           0.000 \\
				& walk 			&     -1.738 &      0.014 &           0.058 &     -1.900 &      0.002 &           0.000 \\
act\_edu 		& auto 			&     -1.357$^*$ &      0.080 &          10.697 &     -0.187 &      0.011 &           0.055 \\
				& walk 			&     -0.067$^*$ &      0.086 &          22.325 &     -0.871 &      0.029 &           0.558 \\
act\_home 		& auto 			&     -0.119$^*$ &      0.026 &           0.151 &      0.340 &      0.003 &           0.001 \\
				& bike 			&     -1.048$^*$ &      0.044 &           3.217 &     -0.705$^*$ &      0.039 &           1.477 \\
				& transit 		&      0.109$^*$ &      0.027 &           0.093 &      0.764 &      0.004 &           0.001 \\
act\_work 		& auto 			&     -0.055$^*$ &      0.027 &           0.115 &      0.276 &      0.003 &           0.003 \\
				& transit 		&     -0.011$^*$ &      0.028 &           0.096 &      0.631 &      0.004 &           0.004 \\
				& auto+transit 	&     -1.853$^*$ &      0.073 &           4.028 &      0.851 &      0.028 &           0.114 \\
act\_meeting 	& bike 			&     -2.776$^*$ &      0.259 &         154.812 &     -1.803$^*$ &      0.174 &         106.564 \\
\midrule
log-likelihood && \multicolumn{3}{l}{-16145} & \multicolumn{3}{l}{-13121} \\
sample size && \multicolumn{3}{l}{42,255} & \multicolumn{3}{l}{42,255} \\
\# of estimated parameters && \multicolumn{3}{l}{138} & \multicolumn{3}{l}{922} \\
max. validation accuracy && \multicolumn{3}{l}{72.01\%} & \multicolumn{3}{l}{76.73\%} \\
\bottomrule
\multicolumn{7}{l}{$^*$: Not statistically significant at p-value $<$ 0.05.}
\end{tabu}
\end{table}

\subsection{Elasticity analysis}
The point aggregate elasticity of $P_n(i)$ with respect to to input $x_n$ is given by the following equation:
\begin{equation}
    E_{x_n}(i) = \frac{dP_n(i)}{dx_n}\frac{x_n}{P(i)}
\end{equation}

The elasticity measures the impact of increasing or decreasing a variable  on the demand of the respective choice. In this case we use \textit{trip\_dist} as the variable and we measure the impact of market share on the \textit{auto}, \textit{bike}, \textit{transit} and \textit{walk} choices. 
Similarly we compute the arc elasticities of $P_n(i)$ with respect to $\hat{x_n}$ when we change the \textit{trip\_dist} by $\Delta x_n$ where $\hat{x_n} = x_n + \Delta x_n$.
\Cref{tab:elasticity} shows the point elasticities obtained from the MNL, MLP and ResLogit model (16 layers).
The ResLogit model show expected signs similar to the MNL model. \textit{Walk} mode show a smaller increase in trip distance than the MNL model, while \textit{Transit} mode shows a more significant impact from trip distance in the ResLogit model compared to the MNL model.
For the MLP model, \textit{transit} mode show a negative sign compared to the MNL model.
Surprisingly, the ResLogit model shows a different sign in \textit{Auto} mode.
Indeed, for \textit{Auto} mode, one should expect negative elasticity.

\begin{figure}[!t]
    \centering
    \includegraphics[width=\textwidth]{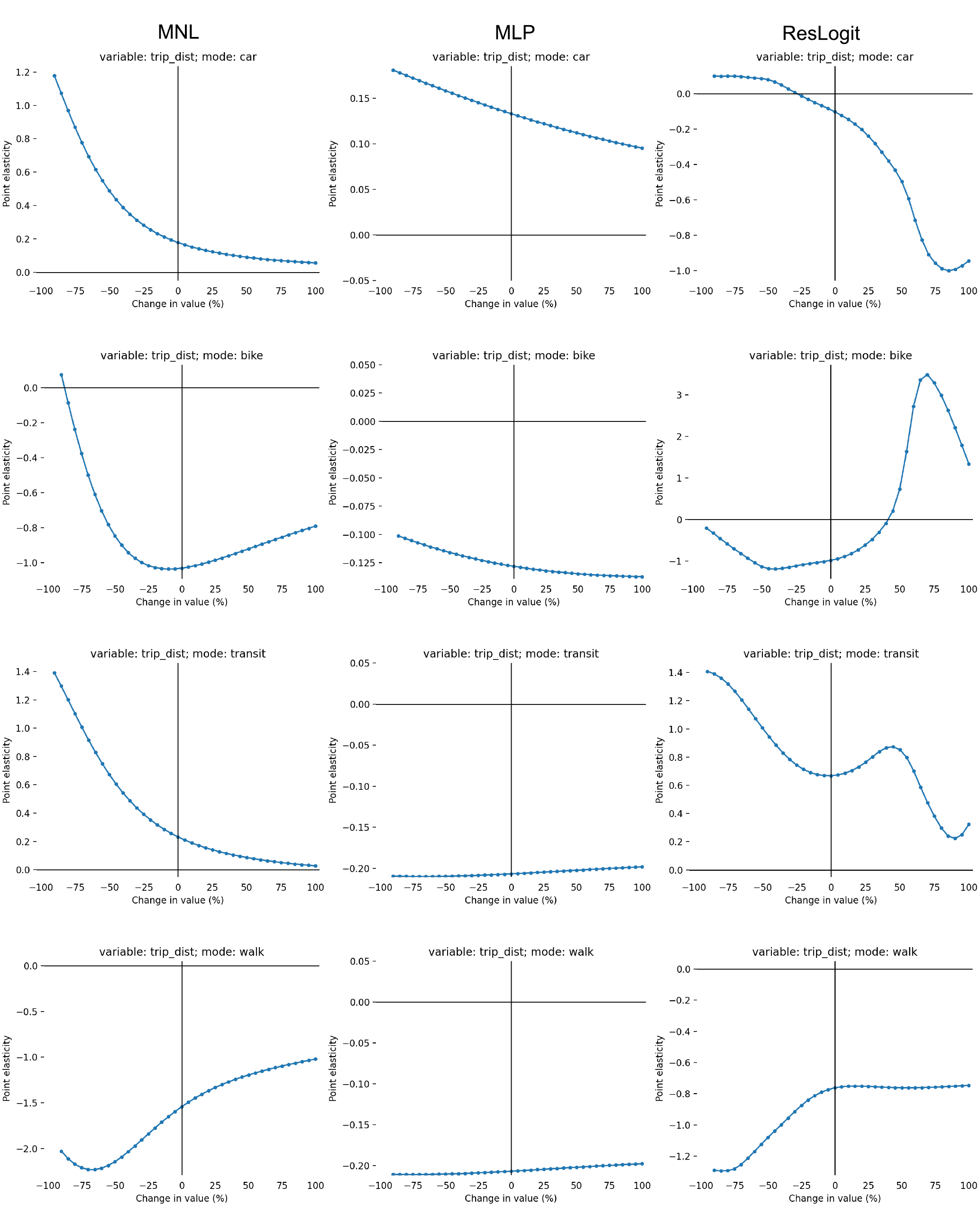}
    \caption{Elasticity versus \% increase or decrease in trip distance, comparison between models.}\label{fig:elasticities}
\end{figure}

If we analyze the two models' elasticities (presented in \Cref{fig:elasticities}) assuming different scenarios where we increase or decrease the overall trip distance, for instance, willingness to change modes to travel a longer or shorter distance or construction of new transit networks. 
We can see that the elasticities from the ResLogit predict a non-linear change relation between trip distance and the respective mode choice. 
This shows a clear distinction from the MLP model, where the relationship between trip distance and mode shows a relatively linear curve.
We expect that elasticity is heterogeneous and it will vary across different scenarios, given different unobserved trade-offs between mode choices, 
For \textit{auto} mode, The ResLogit model predicts that with a decrease in trip distance by 50\%, elasticity is positive (and negative otherwise), while increasing the trip distance will result in greater sensitivity to trip distance.
\textit{Bike} mode shows a positive elasticity when we increase the trip distance by 50\% but a negative elasticity when we decrease the trip distance by about -50\%.
We can infer from this result that travellers are willing to switch from bikes to other modes or from other modes to bikes, considering other unobserved factors not captured in the data.
This sign switching phenomenon is interesting because it indicates a heterogeneous population that will react differently while also \textit{considering other alternatives}.
This consideration of non-chosen alternatives shows that the ResLogit model behaves in line with the behavioural theory of the Mother Logit model example where attributes from non-chosen alternatives enter the utility of the chosen alternative.

\begin{table}[!t]
\centering
\caption{Point elasticities.}
\label{tab:elasticity}
\begin{tabu}{X[1] X[1] X[1] X[1]}
\toprule
 & MNL & MLP & ResLogit \\
Choice & \\
\midrule
&\textit{trip\_dist} & \textit{trip\_dist} & \textit{trip\_dist} \\
Auto & 0.178 & 0.133 & -0.103 \\
Bike & -1.031 & -0.128 & -0.980 \\
Transit & 0.232 & -0.206 & 0.669\\
Walk & -1.54 & -0.207 & -0.769\\
\bottomrule
\end{tabu}
\end{table}

\subsection{Significance of model depth and utility formulation}
The general notion is that increasing the complexity and non-linearity in the model should result in greater model fit, given the higher degree of freedom induced by the neural network.
However, the MLP network model suffers from the vanishing gradient problem shown by increasing the number of layers. 
There is a bottleneck effect with depth $M\geq 4$, and the validation log-likelihood and loss no longer improved.
In contrast, we do not see this detrimental effect in the ResLogit model, even at a depth of 16 layers.
This study highlights how machine learning models may sometimes be worse off than a simple discrete choice model without understanding the neural network formulation structure.

\subsubsection{Behaviour interpretation}
As explained in \Cref{sec:gen_approach}, a decision-maker's learning process may be developed over time through experiences, and the agent updates his or her underlying distribution. 
The ResLogit model captures this effect while retaining the value function of the observed component of the utility.
We can use this approach of capturing uncertainties to account for heterogeneity in the choice process arising from inconsistency within travel mode choice.

Besides the differences in optimal performance, it is also of practical interest to study the actual $\beta$ parameter solution vectors and observe how they differ from a standard MNL model without accounting for learning behaviour.
\Cref{tab:mnl_residual} shows the differences in $\beta$ parameter estimates between the benchmark MNL and RL-16.
These exact set of significant variables accounted for can be inferred from each reported metric's standard error.
A conceptual step in discrete choice analysis is the ability to provide the basis of estimation of $\beta$ (and standard error) and economic indications using data on observed choices and attributes.
Here, our ResLogit approach follows the same approach as discrete choice methods.
The unobserved attributes, expressed in $\varepsilon$ in MNL models, captured the error contribution to the utility.

We observe that the ResLogit counterpart differs from the MNL model in most metrics.
However, the ResLogit model's ability to ``explain away'' uncertainty yields greater parameter significance as reported by the lower standard error.
The formulation of the ResLogit, which adds the $g$ term, captures the cross-effects of the different mode choice alternatives to ensure that the decision is free from unobserved errors and endogeneity.
Under regularity conditions, this residual component captures the unobserved error using a learning algorithm, similar to how in real-life, a traveller explores new route options or stick to habitual choices.
In general, the ResLogit framework allowed for the error term to be formulated within the utility.

\subsubsection{Sensitivity analysis}
It is important to examine the differences in $\beta$ parameter responses when changing the neural network size.
Our emphasis of this analysis is on the $\beta$ value significance and non-linear responsiveness when more residual layers are added to the choice model.

\Cref{tab:variability} shows a sensitivity analysis regarding the $\beta$ parameters of trip time over time of departure. 
The table shows the variation between trip time and time of departure beta parameters for each model. 
The values represent the degree of variability of each time of departure dummy variable on the utility of each mode alternative.
We take the ratio of $(\beta_{\textrm{trip time}}x_{\textrm{trip time}}) / (\beta_{\textrm{departure dummy}}x_{\textrm{departure dummy}})$.
This gives us the sensitivity of travel time over different departure time segments.
If the parameters for $\beta_{\textrm{trip time}}$ are not influenced by variation in departure time, then the values would have a small standard deviation across departure time, and the standard deviation would give an indicator of uniformity of the trip time-sensitivity across different departure times.
If the standard deviation is small, it would indicate that the trip time heterogeneity is captured in the residual component. The $\beta_{\textrm{trip time}}$ represents the value that is closer to the true mean.
The attribute effects are shown in \Cref{tab:variability} represent the mean preference on each individual's utility, after controlling for taste variability.
This result indicates the effects of increasing residual layers on the stability of the econometric parameters.

As expected in the MNL model, the time of departure dummy variable influences the utility and choice of mode.
This is a consistent result, as we cannot represent the variation over the departure time as a single linear factor in the utility function.
Modifying the MNL model by incorporating the residual layers would reduce the variability and sensitivity to time of departure.
The average standard deviation of trip time versus time of departure coefficient decreases as we increase the number of layers is shown in the table.
This shows how the implied heterogeneity in the utility function can be explained away through the neural network component, retaining the properties of the observed utility component.
Note that this estimation does not allow us to identify the relationship between the heterogeneity of departure time and the preference of different travel modes.
One can use economic indicators to estimate this effect.
The values reported for RL-2 to RL-16 may not be entirely stable, and further investigation is needed into how the model responds to changes in hyperparameters and regularization.
However, we can conclude that this experiment shows the capability of our proposed ResLogit approach, particularly in: (a) allowing for a specific analysis of the underlying distribution and (b) exploring the attributes that represent the most significant degree of heterogeneity in the model--that may present an interesting subject for future research.

\begin{table}[!t]
\centering
\caption{Sensitivity analysis of different travel modes over time of departure. Values show the difference in trip time parameter estimates across hourly segments.}
\label{tab:variability}
\resizebox{0.74\textwidth}{!}{
\begin{tabu}{X[1,l] X[0.5,l] X[0.5,l] X[0.5,l] X[0.5,l]}
\toprule									
Model & \multicolumn{4}{l}{trip time/time of departure variability} \\
\midrule									
MNL & Auto & Bike & PT & Walk\\
hour\_8\_10 & 3.07 & -0.26 & 26.36 & -1.03 \\
hour\_11\_13 & -3.34 & -0.24 & -12.08 & -5.43 \\
hour\_14\_16 & -2.07 & -0.33 & 5.88 & -2.61 \\
hour\_17\_19 & -10.81 & -0.45 & 3.26 & -1.75 \\
hour\_20\_22 & 2.42 & -0.16 & -3.49 & -1.31 \\
hour\_23\_1 & 0.76 & -0.14 & -1.26 & -0.54 \\
hour\_2\_4 & 1.88 & -0.09 & -0.52 & -0.43 \\
hour\_5\_7 & 4.86 & -0.16 & -14.26 & -0.92 \\\\[-0.75em]
stddev & 4.66 & 0.11 & 11.74 & 1.54 \\
\midrule
RL-2 \\
hour\_8\_10 & -0.18 & 3.63 & 15.37 & -0.60 \\
hour\_11\_13 & -0.16 & 1.73 & -5.99 & -1.33 \\
hour\_14\_16 & -0.17 & 2.19 & -14.66 & -0.78 \\
hour\_17\_19 & -0.23 & 3.76 & 9.25 & -0.54 \\
hour\_20\_22 & -0.19 & 1.77 & -6.62 & -0.83 \\
hour\_23\_1 & -0.29 & 1.65 & -2.07 & -0.64 \\
hour\_2\_4 & -0.17 & 1.68 & -0.89 & -0.94 \\
hour\_5\_7 & -0.17 & 2.29 & 45.38 & -0.72 \\\\[-0.75em]
stddev & 0.04 & 0.81 & 17.62 & 0.23 \\
\midrule
RL-4 \\
hour\_8\_10 & 0.08 & 0.19 & 0.84 & -1.24 \\
hour\_11\_13 & 0.06 & 0.23 & 1.01 & -1.80 \\
hour\_14\_16 & 0.08 & 0.22 & 1.07 & -1.47 \\
hour\_17\_19 & 0.10 & 0.24 & 1.08 & -1.58 \\
hour\_20\_22 & 0.08 & 0.19 & 0.97 & -1.49 \\
hour\_23\_1 & 0.09 & 0.16 & 0.94 & -1.09 \\
hour\_2\_4 & 0.07 & 0.20 & 1.88 & -1.50 \\
hour\_5\_7 & 0.08 & 0.22 & 0.88 & -1.66 \\\\[-0.75em]
stddev & 0.01 & 0.03 & 0.31 & 0.21 \\
\midrule
RL-8 \\
hour\_8\_10 & -0.26 & -1.35 & 0.39 & -1.39 \\
hour\_11\_13 & -0.26 & -2.89 & 1.42 & -1.15 \\
hour\_14\_16 & -0.24 & -1.59 & 0.48 & -1.36 \\
hour\_17\_19 & -0.27 & -1.58 & 0.41 & -1.61 \\
hour\_20\_22 & -0.25 & -2.01 & 0.45 & -1.85 \\
hour\_23\_1 & -0.37 & -1.46 & 0.34 & -1.12 \\
hour\_2\_4 & -0.26 & 3.87 & -2.94 & -0.75 \\
hour\_5\_7 & -0.25 & -2.14 & 0.38 & -1.53 \\\\[-0.75em]
stddev & 0.04 & 1.95 & 1.20 & 0.32 \\
\midrule
RL-16 \\
hour\_8\_10 & 0.75 & -0.93 & 0.40 & 0.52 \\
hour\_11\_13 & 0.69 & -0.67 & 0.49 & 0.42 \\
hour\_14\_16 & 0.71 & -0.81 & 0.43 & 0.46 \\
hour\_17\_19 & 0.68 & -1.22 & 0.40 & 0.48 \\
hour\_20\_22 & 0.70 & -0.86 & 0.39 & 0.50 \\
hour\_23\_1 & 0.62 & -0.87 & 0.43 & 0.55 \\
hour\_2\_4 & 0.51 & -0.56 & 1.04 & 0.48 \\
hour\_5\_7 & 0.69 & -0.91 & 0.42 & 0.50 \\\\[-0.75em]
stddev & 0.07 & 0.18 & 0.21 & 0.04 \\
\bottomrule
\end{tabu}
}
\end{table}

	\section{Conclusion}
\label{sec:conclusion}
This paper has presented a data-driven deep learning-based choice model that integrates a residual neural network architecture into a Logit model structure.
This paper's methodological contribution is a new model that captures the learning process using neural network model structure for accounting for cross-effects in the utility error term.
We proposed an approach that combines a residual neural network with a Logit model.
This study's first objective resolves the shortcomings in the integration of machine learning techniques and neural networks in discrete choice modelling.
The second objective addressed the systematic error of biased model estimates in DNNs due to its lack of economic interpretability.

Unlike earlier studies that only examined the performance of machine learning algorithms and their comparisons with discrete choice models in out-of-sample predictions, this paper studies the impact of a residual function in the choice utility as a data-driven variant of the Mother Logit model.
The ResLogit model proposed in this paper frames the Mother Logit model's expansion function like a neural network and the parameters within the neural network are estimated through a mini-batch stochastic gradient descent algorithm, maximizing over the out-of-sample validation set.
This data-driven approach also addresses model non-identifiability issues when estimating a large number of unknown parameters.
A new direction to a more flexible and general model is presented using the concept of residual modelling -- mapping the error term correlation to a residual function instead of using traditional neural networks.
The skipped connection structure allows each residual layer to be estimated independently without model identification problems due to exploding/vanishing gradient during backpropagation.

A classic red/blue bus IIA violation example is used, and we demonstrated our methodology on a large scale travel behaviour dataset.
We examined the performance comparison with an MNL and MLP neural network across a different number of layers.
The results showed that with a ResLogit model, it optimized quickly and efficiently, without degradation in model performance as the number of layers increased.
In the context of model identifiability, the ResLogit model yielded a smaller standard error for each econometric model parameters than the baseline MNL model.
We also demonstrated the sensitivity of trip time and time of departure variability over different model characteristics. We observed that incorporating residual layers reduced model sensitivity to cross-effects and choice heterogeneity.

Our proposed ResLogit model improved discrete choice models' capabilities in terms of performance without sacrificing model interpretability.
We noted that our experiment results do not consider hyperparameter tuning or regularization steps, which may affect the reliability of our model validation results.
This proof-of-concept illustrates how choice modellers can leverage on deep learning methodologies and learning algorithms to enhance the current set of tools and models for discrete choice analysis.
Our future work will establish additional models and extensions to our proposed ResLogit methodology.

More work has to be done on the interpretability of the model, and how to define clear guidelines so researchers without advanced knowledge of machine learning can use these new modelling techniques.
Also, more comparative studies can be done between different learning algorithms for Logit models.
Further investigation is also required into the meta-learning side of deep learning in discrete choice modelling.
For example, we do not know the optimal hyperparameter configuration or efficiently identify a good set of hyperparameters without a tedious iterative search.

	\clearpage
	
	\bibliographystyle{elsarticle-harv}

	\bibliography{bibliography}

\end{document}